\documentclass[useAMS,usenatbib]{mn2e}
\usepackage{graphicx}
\usepackage{times}
\usepackage{color}

\title[Periodicities in RX J0146.9+6121]{Periodicities in the high-mass X-ray binary system RX J0146.9+6121/LS I+61$^{\circ}$235}
\author[G.E. Sarty et al.]{Gordon E. Sarty$^{1}$\thanks{E-mail:
gordon.sarty@usask.ca}, L\'{a}szl\'{o} L. Kiss$^{2}$,
Richard Huziak$^{1}$, Lionel J.J. Catalan$^{3}$, \vspace*{0.4em}  \\ \LARGE \rm Diane Luciuk$^{1}$, 
Timothy R. Crawford$^{4}$,
David J. Lane$^{5}$, Roger D. Pickard$^{4}$,  \vspace*{0.4em} \\
\LARGE \rm Thomas A. Grzybowski$^{4}$, Pere Closas$^{4}$, Helen Johnston$^{2}$, David Balam$^{6}$
and \vspace*{0.4em} \\ \LARGE \rm Kinwah Wu$^{7}$ \\
$^{1}$Department of Physics and Engineering Physics, University of Saskatchewan, 9 Campus Drive, Saskatoon, Saskatchewan S7N 5A5, Canada \\
$^{2}$Institute of Astronomy, School of Physics A28, University of Sydney, New South Wales 2006, Australia\\
$^{3}$Lakehead University, 955 Oliver Rd., Thunder Bay, Ontario P7B 5E1, Canada\\
$^{4}$American Association of Variable Star Observers, 49 Bay State Rd., Cambridge, MA 02138, USA\\
$^{5}$Department of Astronomy and Physics, Saint Mary's University, 923 Robie St., Halifax, Nova Scotia
B3H 3C3, Canada \\
$^{6}$Department of Physics and Astronomy, University of Victoria, PO Box 3055, STN CSC, Victoria, British Columbia, V8W 3P6 Canada \\
$^{7}$Mullard Space Science Laboratory, University College London, 
Holmbury St.~Mary, Dorking, Surrey RH5 6NT, United Kingdom
}

\begin{document}

\date{in original form August 2008}

\pagerange{\pageref{firstpage}--\pageref{lastpage}} \pubyear{2008}

\maketitle

\label{firstpage}

\begin{abstract}
The high-mass X-ray binary RX J0146.9+6121, with optical counterpart LS I+61$^{\circ}$235 (V831 Cas),
is an intriguing system on the outskirts of the open cluster NGC 663. It contains
the slowest X-ray pulsar known with a pulse period of around 1400s and,
primarily from the study of variation in the emission line profile of H$\alpha$, it is known
to have a Be decretion disk with a one-armed density wave period of approximately 1240d. Here we present the results of an extensive photometric campaign, supplemented with optical spectroscopy, aimed at measuring short time-scale periodicities.
We find three significant periodicities in the photometric data at, in order of statistical significance, 0.34d, 0.67d and 0.10d. 
We give arguments to support the interpretation that the 0.34d and 0.10d periods could be due to stellar oscillations of the B type primary star and that the 0.67d period is the spin period of the Be star with a spin axis inclination of 23$^{+10}_{-8}$ degrees. We measured a systemic velocity of $-37.0 \pm 4.3$ km s$^{-1}$ confirming that LS I+61$^{\circ}$235 has a high probability of membership in the young cluster NGC 663 from which the system's age can be estimated as 20--25 Myr. From archival 
{\it RXTE} ASM  data we further find ``super'' X-ray outbursts roughly every 450d. If these super outbursts are caused by the alignment of the compact star with the one-armed decretion disk enhancement, then the orbital period is approximately 330d.
\end{abstract}


\section{Introduction}\label{sec1}

High-Mass X-ray Binaries (HMXBs) are interacting binary stars which emit X-rays as a result of mass transfer and accretion processes. Be X-ray binaries
(BeXs) consist of a rapidly rotating main sequence Be star with an equatorial decretion disk, which gives rise to the optical emission lines, orbited by a compact object that is usually a neutron star. The other major class of HMXBs is composed of the supergiant X-ray binaries (SGXs) that consist of a more evolved OB star closely orbited by a neutron star (or in some cases a black hole, like Cyg X-1).
SGXs tend to have short (a few days) orbital periods while the orbital periods of BeXs are considerably longer.
BeXs are generally transient X-ray sources with outbursts occurring when the compact star passes periastron 
\citep{okazaki2001}. All HMXBs are relatively young, rapidly evolving systems in which one component, in most known cases, has already evolved to a supernova and subsequently to the current compact object. Mass redistribution within the system together with mass loss may prevent the supernova, however, and leave a white dwarf as the compact object. The HMXB $\gamma$ Cas, for example, is suspected of containing a white dwarf \citep{lopesdeoliveria2006,haberl1995}. Since a significant fraction of all stars form as binary stars, a complete understanding of the evolution of high-mass stars requires an understanding of their evolution in the presence of mass transfer in the interacting binary situation.  

The physical properties of many HMXB systems, in particular their orbital period, are known from their X-ray behaviour. But the orbital periods of many Galactic systems (approximately half of those with known optical counterparts) remain unknown \citep{liu2006}. We have recently initiated a long-term program to obtain time-series optical photometric data for bright HMXBs in an effort to measure periodicities and determine the orbital periods \citep{sarty2007}. Here we report the results of one of our first observing campaigns aimed at RX J0146.9+6121/LS I+61$^{\circ}$235.
The data were obtained by observers associated with the American Association of Variable Star Observers (AAVSO) in response to our call to observe HMXBs on an on-going basis. In addition to the photometry, we also made optical spectroscopic radial velocity measurements using the
Dominion Astrophysical Observatory's Plaskett telescope. Significant photometric and radial velocity variation was found which we report and interpret here.
We further interpret the optical data alongside publicly available X-ray light curves from the
All Sky Monitor (ASM) on board the Rossi X-ray Timing Explorer ({\it RXTE}) satellite  
and previously published information to make an estimate of the orbital period. This paper is organised as follows. The optical and X-ray observations are described in Sect.~\ref{sec2}. The results of period searches in the optical and X-ray light curves are
presented in Sect.~\ref{sec3}. Interpretation of the results is given in Sect.~\ref{sec4}, where we present evidence for $\beta$ Cephei type pulsations in the optical component of the system, confirm its membership in the star cluster NGC 663 and argue that the orbital period is about 330d. We conclude the paper in Sect.~\ref{sec5}.

\section{Observations}\label{sec2}

\subsection{Optical photometry}

\begin{figure}
\vspace*{-5em}
\includegraphics[scale=0.4]{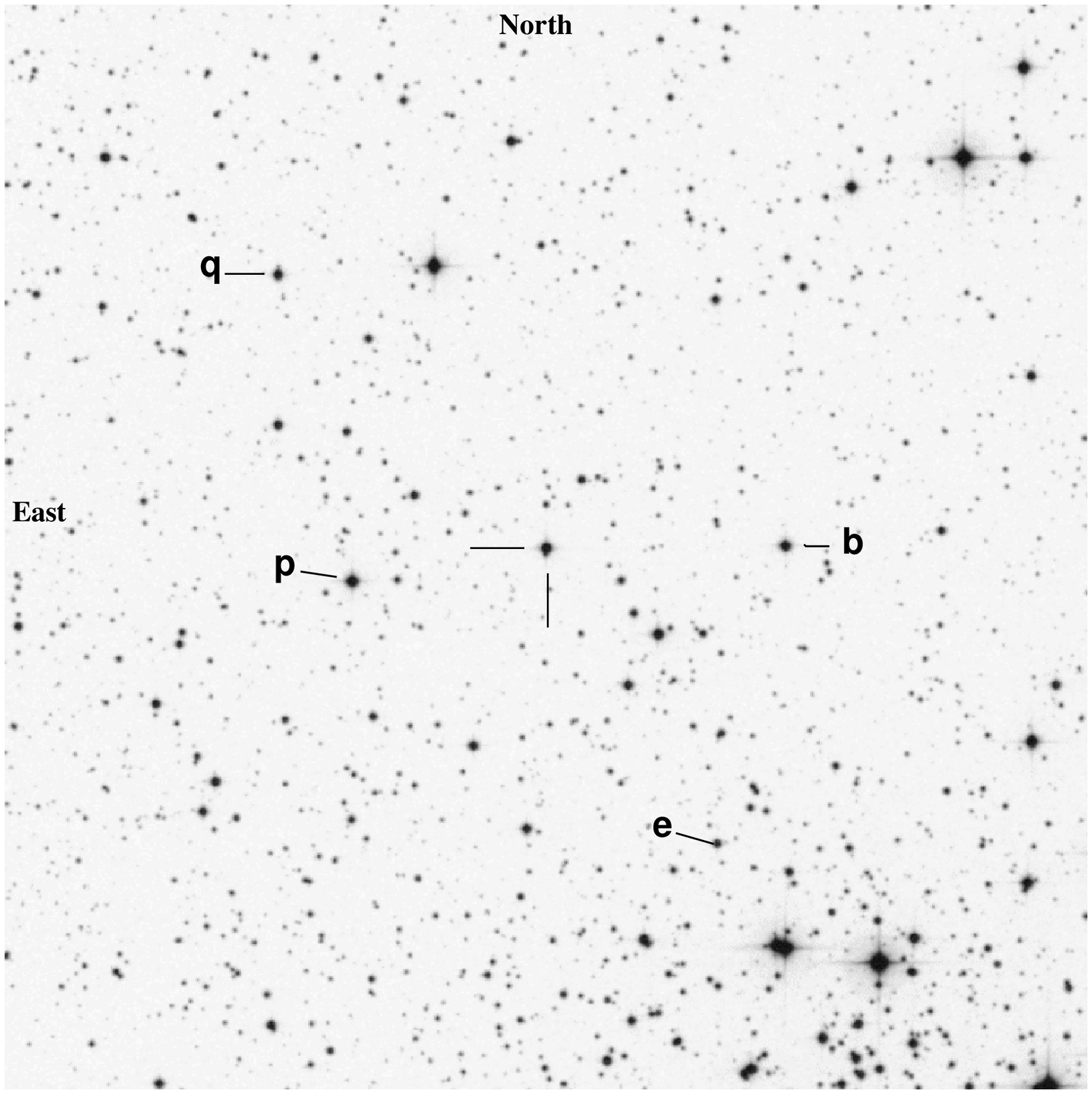}
\vspace*{-7em}
\begin{center}
\begin{tabular}{cccccc}
\multicolumn{6}{c}{Comparison/Check stars} \\
\hline
Star & RA           & Dec            &  $B$       & $V$         &   $I_{C}$      \\
b   & 01 46 32.64   &   61  21  18.0 & 12.464	& 11.684    &   10.643  \\	
e	& 01 46	41.76	&	61	17	13.2 &	13.560	&	12.976	&	12.183 	 \\	
p	& 01 47	22.80	&	61	21	03.6 &	11.378	&	10.857	&	10.124	\\				
q	& 01 47	30.00	&	61	25	19.2 &	12.625	&	12.041	&	11.334

\end{tabular}
\end{center}
\caption{Comparison stars used by the AAVSO observers for the differential photometry
reported here. The target star, RX J0146.9+6121/LS I+61$^{\circ}$235, is marked by the two dashes. The data are based on measurements made at the Sonoita Research Observatory (SRO)
by Arne Henden and the constancy of the stars has been verified from data obtained by observers HUZ and CTE. This image is based on the POSS2/UKSTU Red images from the STScI Digitized Sky Survey. The
field of view is 0.25$^{\circ}$.
\label{figcomps}}
\end{figure}

\begin{figure*}
\includegraphics[scale=1]{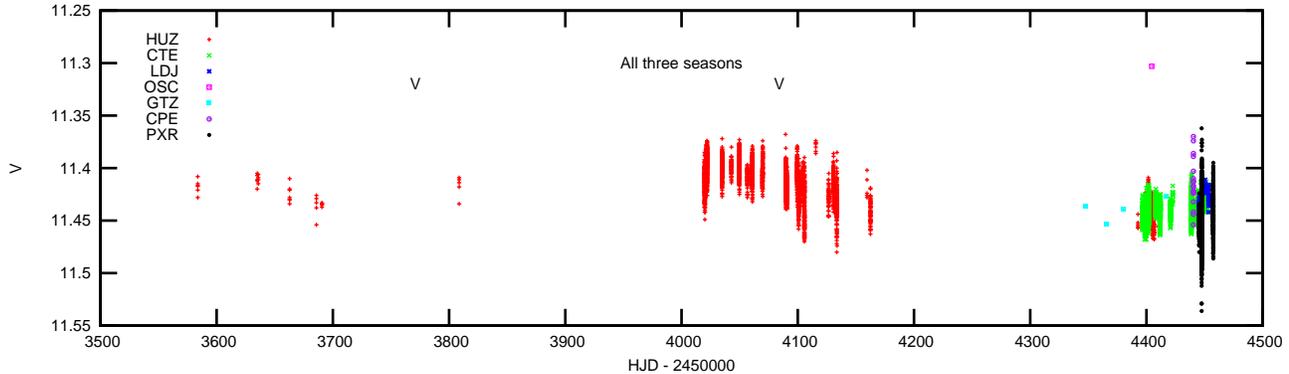}
\caption{The complete $V$ band light curve. The first season covers HJD 2453583 to 2453808, the second season covers HJD 2454019 to 2454162, and the third season covers HJD 2454347 to 2454457. No long term changes greater than about 0.05 magnitude are seen over the approximately 1000 days of observations. 
The two ``V'' symbols mark the times of the two largest X-ray outbursts during the three observing seasons as seen by the {\em RXTE} ASM and shown in Fig.~\ref{rxte}(a).
\label{fig1}}
\end{figure*}

\begin{figure}
\hspace*{-0.9em}
\includegraphics[scale=1]{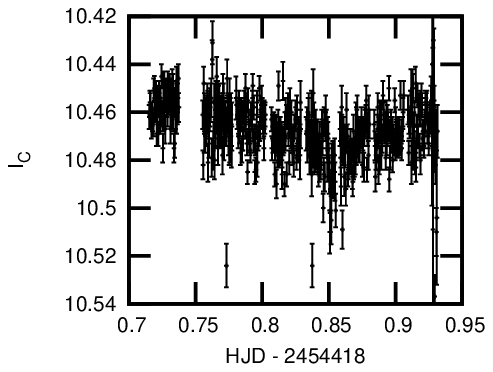}
\hspace*{-3em}
\includegraphics[scale=1]{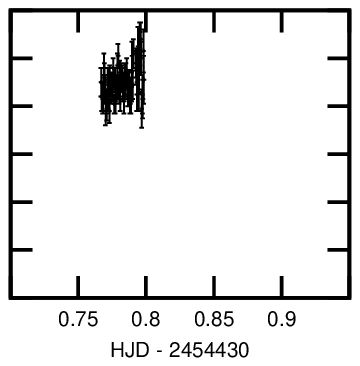}
\caption{Cousins $I_{C}$ band photometric data provided by observer CTX. \label{figIB-V}}
\end{figure}

Three observing seasons of optical photometric data have been obtained of LS I+61$^{\circ}$235 covering the
ends of 2005, 2006 and 2007. These data were obtained primarily in the Johnson $V$ band by the observers
listed in Table \ref{tab1}, where they are identified by the observer initials assigned to them by the
AAVSO. The complete $V$ band dataset is shown in Fig.~\ref{fig1}, where it may be seen that the first two seasons
were covered exclusively by HUZ. Cousins $I_{C}$ band data are shown in Fig.~\ref{figIB-V} and a limited set of Johnson $B-V$ colour observations were obtained by LDJ. Other observers joined the HMXB observation program in the third season
in response to a general call for HMXB observations \citep{sarty2007} and to AAVSO Alert Notices 348 and
354. All the photometric observations are available from the AAVSO International Database. The total numbers
of observations are shown in Table \ref{totals}. The observations were initially retrieved from the
AAVSO International Database and then the observers were individually contacted to resolve discrepancies and
errors, if any. This ``validation'' step is necessary to restore some homogeneity to an otherwise relatively inhomogeneous dataset. All observers reported an estimate of the signal to noise ratio (S/N) of each
measurement, usually based on a formal estimate of the Poisson shot noise. All of the photometry reported was differential photometry with the magnitudes being determined relative to the comparison stars labelled in Fig.~\ref{figcomps}. A variety of methods were used by the observers to reduce the data using aperture photometry and software like
MaxImDL, Aip4Win or MiraAP. None of the data were transformed to the standard Johnson/Cousins photometric system and therefore are instrumental magnitudes. Since our main aim is to determine periodicities from the light curve, instrumental magnitudes are as valuable as properly standardised data provided there are no large unaccounted offsets between the subsets. The majority of our data were obtained independently by two observers. The photometric values reported by those two observers, at times of simultaneous observation, show excellent agreement.

\begin{table}
\caption{List of observers and their equipment.
\label{tab1}}
\begin{tabular}{lll}
\hline
Observer                  & Telescope              & CCD \\
\hline
HUZ, R Huziak 	   & 30-cm SCT              & SBIG ST-9XE \\
                   & 30-cm SCT              & STL-1301E \\
CTE, L Catalan 	   & 25-cm Newtonian              & KAI-2020M \\
LDJ, DJ Lane       & 28-cm SCT              & SBIG ST-9XE \\
CTX, TR Crawford   & 30-cm SCT              & SBIG ST-9XE \\
PXR, RD Pickard    & 30-cm SCT              & Starlight Xpress SXV-H9 \\
CPE, P Closas 	   & 15-cm refractor        & Starlight HX516 \\
OSC, SC Orlando    & 20-cm SCT              & SBIG ST-8XE \\
GTZ, TA Grzybowski & 25-cm SCT              & SBIG ST-402 \\
\hline
\end{tabular}
\end{table}

\begin{table}
\caption{List of total numbers of photometric observations by the AAVSO observers broken
down by filter. 
\label{totals}}
\begin{center}
\begin{tabular}{cccccc}
\hline
Observer   & $B$    & $V$    &  $I_{C}$  & Total\\
\hline
HUZ 	   &      & 4702 &           & 4702 \\
CTE 	   &      & 3961 &           & 3961 \\
LDJ        &  64  & 66   &           & 130 \\
CTX        &      &      & 556       & 556 \\
PXR        &      & 1484 &           & 1484 \\
CPE 	   &      & 18   &           & 18 \\
OSC        &  1   & 1    &           & 2 \\
GTZ        &      & 4    &           & 4 \\
Totals     &  65  & 10236 & 556      &  10857 \\
\hline
\end{tabular}
\end{center}
\end{table}

\subsection{Optical spectroscopy}

Optical spectroscopy at a resolution of approximately 0.4 \AA\ (10 \AA\ /mm, $\lambda/\Delta \lambda \sim$ 11000)
was obtained during two observing runs with the Cassegrain spectrograph on the 1.85-m Plaskett telescope
at the Dominion Astrophysical Observatory (DAO) in Victoria, British Columbia, Canada. The two observing runs were from August 20 to September 3, 2007 and from December 3 to December 12, 2007. The spectrograph was nominally set to receive the band from 4300 to 4550 \AA\ with a 0.005-in slit in the ``21181'' configuration, meaning that a 21-in focal length camera, 18-hundred grooves per mm grating and the 1st spectral order were used. The
CCD was a SITe-2, with 1752$\times$532 15 $\mu$m pixels. Exposure times varied from 900 to 2400s. The spectra were extracted and wavelength 
calibrated against the emission lines from an Fe-Ar lamp using the IRAF {\tt doslit} task.

\begin{figure*}
\includegraphics[scale=1]{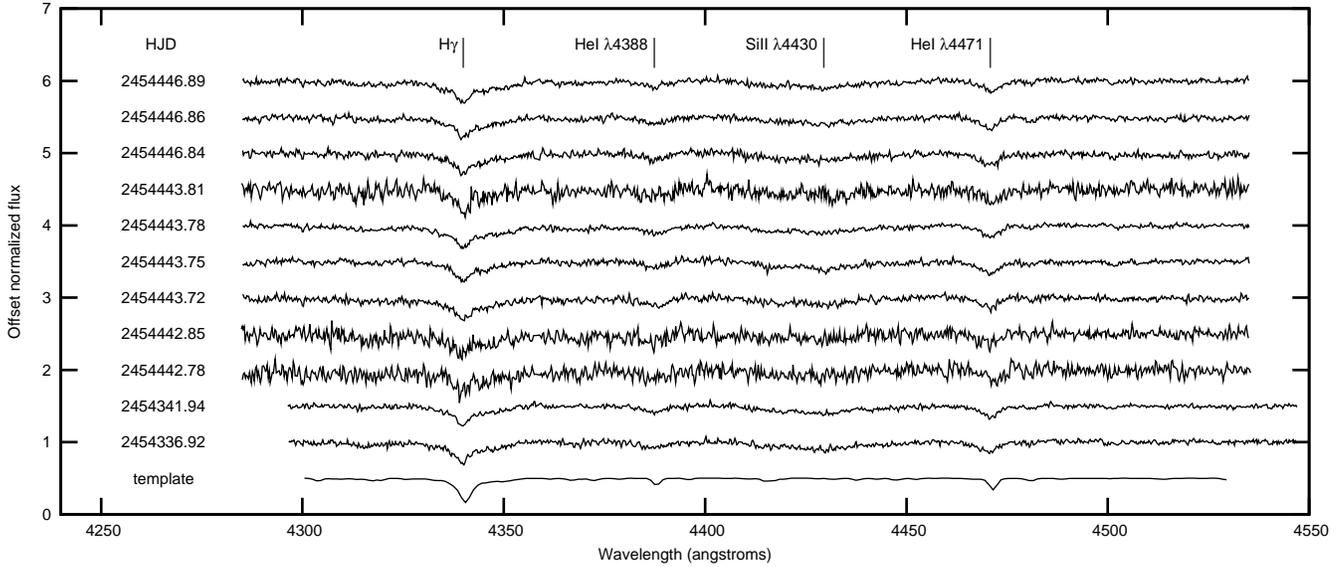}
\caption{Spectra of LS I+61$^{\circ}$235 obtained from two DAO observing runs. The spectra are dominated
by four lines: H$\gamma$ ($\lambda$4340), SiII ($\lambda$4430) and
He I ($\lambda\lambda$4388, 4471).\label{fig2}}
\end{figure*}

A total of 11 spectra were obtained as shown in Fig.~\ref{fig2}. Of those spectra, 3 were quite noisy because of passing clouds. Radial velocities were measured from these spectra as described in Sect.~\ref{sec3}.

\subsection{RXTE ASM X-ray data}

\begin{figure}
\includegraphics[scale=1]{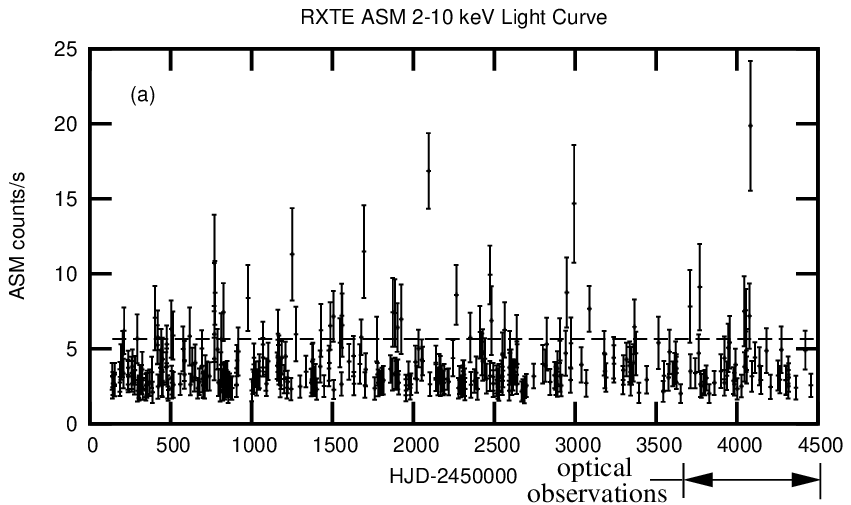}
\includegraphics[scale=1]{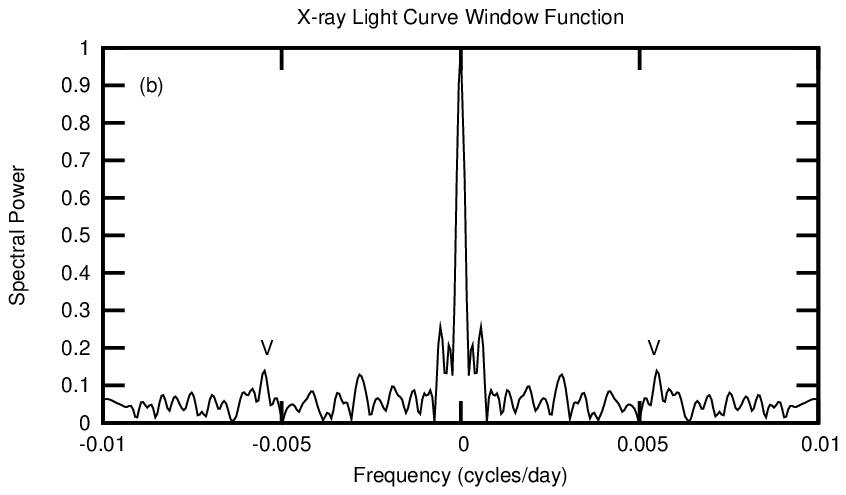}
\caption{(a) {\it RXTE} ASM 2-10 keV X-ray light curve. The dashed line shows the mean background level, which is essentially subtracted from the three-sigma detections, and the time span of our photometric observations is shown. The major outbursts are
separated by about 450d with smaller outbursts occurring approximately every 90d.
(b) The window function for the sampling pattern underlying the ASM light curve. There is no large
concentration of spectral power in the window function at a period of 
184d (0.0054 cycles/day)
although a small local maximum occurs there 
(marked with the ``V'' symbols).
\label{rxte}}
\end{figure}

Archived data from the {\it RXTE} ASM were used to identify outbursts that occurred before and
during our optical observing runs. The {\it RXTE} ASM is composed of three Scanning Shadow Cameras (SSCs) that perform sets of 90s pointed
observations (dwells) covering about 80\% of the sky every $\sim$90 minutes \citep{levine1996}.
Definitive ASM data for RX J0146.9+6121 were downloaded from the {\it RXTE} Guest Observer Facility.
The data used were from the dwell by dwell compilation. Each raw data point represents the fitted source flux from one 90 second
dwell. Data from all three SSCs were used and represent nominal 2-10 keV rates in ASM counts/second (c/s).
Nominally, the Crab nebula flux is about 75 ASM c/s (when the source is at the centre of an SSC field of view and all 8 anodes are operational). We retained only the ``three-sigma'' detections based on the reported variance for each data point. The ASM data are shown in Fig.~\ref{rxte}.

\section{Results}\label{sec3}

\subsection{X-ray outbursts}

The previously reported outburst of 1997 \citep{mereghetti2000,reig1999} can be seen in the X-ray light curve at HJD 2450769.
Subsequent outbursts are easily identified, with the largest outburst occurring at HJD 2454083. The large outburst at HJD 2454083 (see Fig.~\ref{rxte})
corresponds to a possible brightening of the $V$ band light curve in the second observing season peaking at approximately HJD 2454050 (2006, see Fig.~\ref{fig1}) and appears to
be the peak of X-ray activity lasting about 40 days. The peak of the 2006 $V$ band brightening appears to precede, by a few tens of days, the
X-ray peak. The peak
in $V$ band brightness seen at about HJD 2453630 in the first observing season (2005) precedes, by roughly 80 days, the X-ray outburst marked by X-ray flux of about 7.5 ASM c/s
($\sim$0.1 Crab) at HJD 2453710 and 2453770. In this case the $V$ brightness appears to have faded considerably (by about 0.03 mag) by the time the X-ray outburst occurs. 

A formal search for periods in the ASM data was done using the {\tt period04} \citep{lenz2005} software,
a Lomb-Scargle periodogram \citep{lomb1976,scargle1982} and a phase dispersion minimisation (pdm) technique \citep{stellingwerf1978}. All methods report a strong 184d period although the third subharmonic of that period (552d) is stronger in the pdm analysis. The window function, representing the Fourier transform of the light curve sampling pattern, shows no structure (strong peaks) that could cause a 184d period to appear in the periodograms (Fig.~\ref{rxte}(b)). However, folding the X-ray light curve at 184d reveals no obvious pattern. A visual inspection of the ASM light curve reveals an outburst roughly every 90d with every fifth outburst (every 450d) being of much larger amplitude.

\subsection{$I_{C}$ band and $B-V$ colour}

Referring to Fig.~\ref{figIB-V}, the $I_{C}$ band data shows a light curve shape very similar to that seen in the $V$ band data (see below). 
At this point however, we cannot be certain that the $I_{C}$ light curve varies with the same frequency and phase as the
$V$ light curve. Simultaneous (or nearly so) observations in $I_{C}$ and $V$ are required to assess the relationship between the
two light curves or, equivalently, if there are changes in the $V-I$ colour. 

Three nights of $B-V$ data from observer LDJ show signs of variability, indicating that the amplitude and/or frequency and/or phase of the $B$ and $V$ light curves are
different. Again more multi-colour photometry is required before anything positive can be claimed about colour variations.

\subsection{Photometric period analysis of the $V$ band data}

\begin{figure*}
\includegraphics[scale=1]{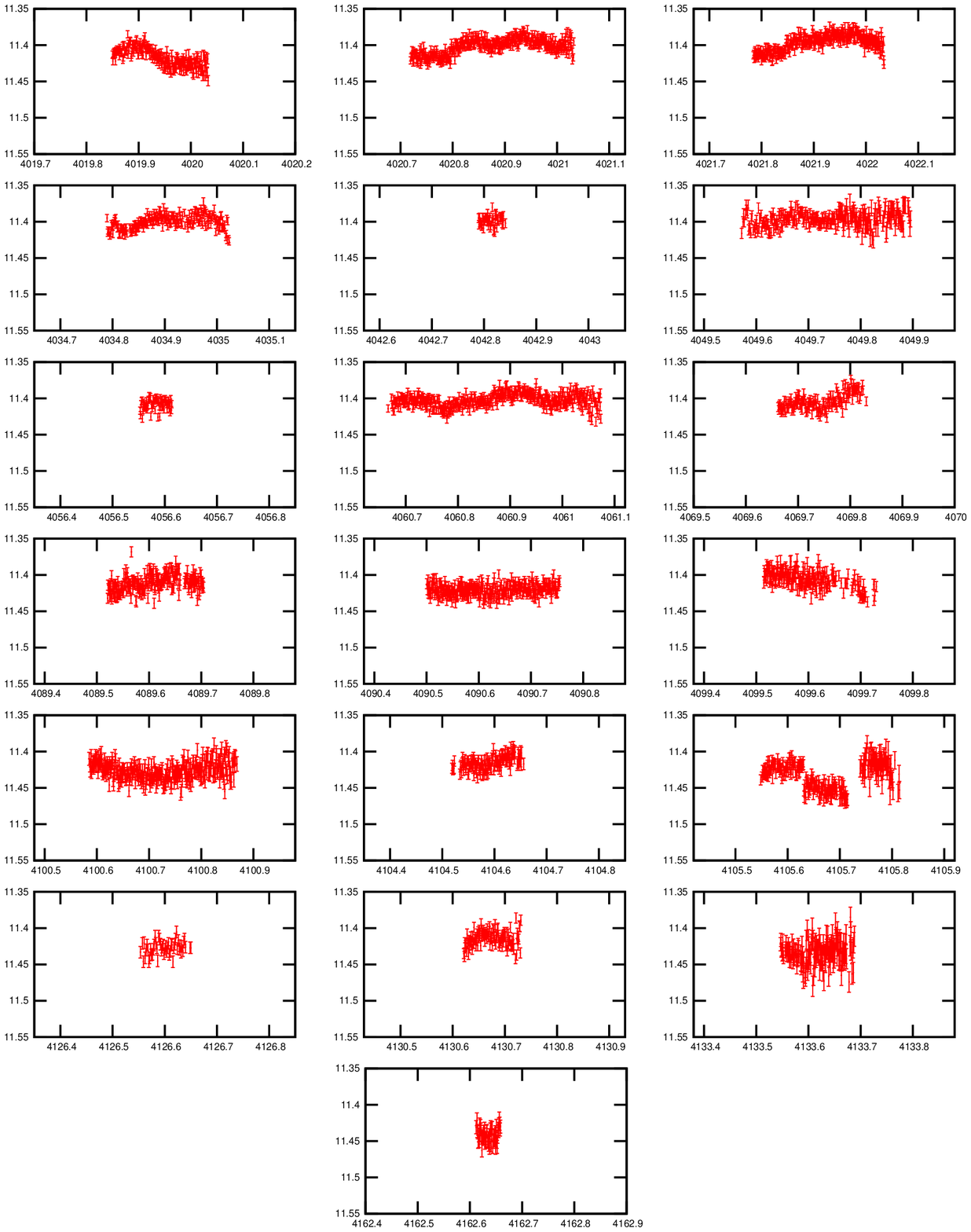}
\caption{Night by night presentation of season 2 $V$ band data. Abscissas are HJD$-$2450000, ordinates are instrumental $V$ magnitude. Observer symbols are as in Fig.~\ref{fig1}; in this case HUZ is the only observer. Some nights where only a few data points were taken are not shown. Note the presence of short period fluctuations ($\sim$0.1d) on every night.
\label{fig1a}}
\end{figure*}

\begin{figure*}
\includegraphics[scale=1]{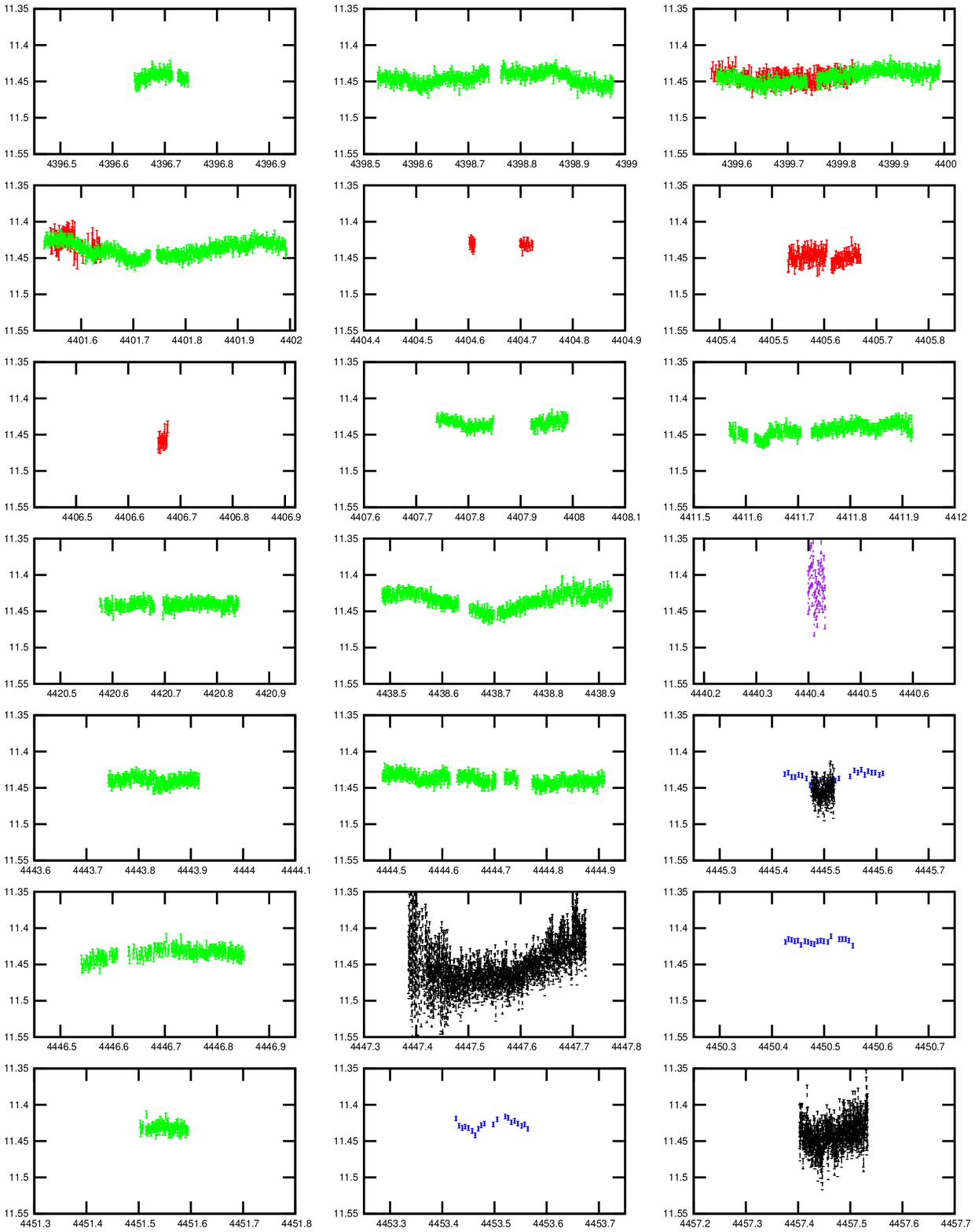}
\caption{Night by night presentation of season 3 $V$ band data. Abscissas are HJD$-$2450000, ordinates are instrumental $V$ magnitude. Observer symbols are as in Fig.~\ref{fig1}. Some nights where only a few data points were taken are not shown.
\label{fig1b}}
\end{figure*}

Iterative Fourier analysis of the $V$ band data, shown in detail in Figs.~\ref{fig1a} and \ref{fig1b} for the second and third seasons, was done using the software {\tt period04} \citep{lenz2005} in which
the data were prewhitened with the dominant frequency between iterations. Only the data from observers HUZ and CTE were included in the Fourier analysis because these data consistently had millimag precision and were compatible without transformation to a standard photometric system. The analysis was repeated three times; once for the whole dataset (840 days),
once for the second season (160 days), and once for the last part of the last season (30 days). In addition, each of the
three analyses was done twice, once without weighting the data and a second time weighting the data with
weights $w=1/\sigma^{2}$ where $\sigma$ is the reported error. The frequencies found from the iterative weighted Fourier analysis are shown in Table \ref{pertab}.

\begin{figure}
\includegraphics[scale=1,angle=-90]{fig8a}
\includegraphics[scale=1,angle=-90]{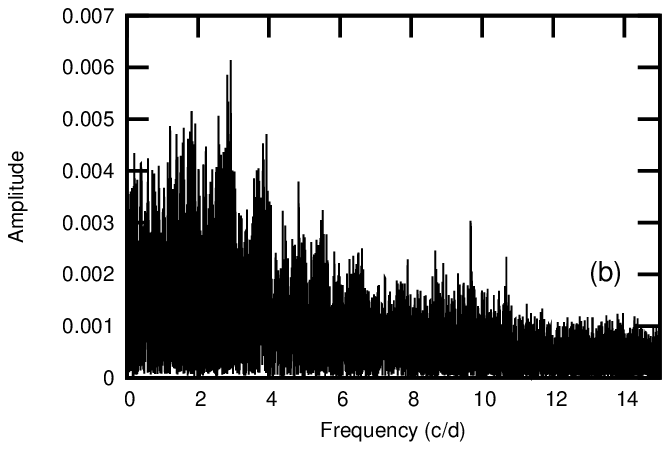}
\includegraphics[scale=1,angle=-90]{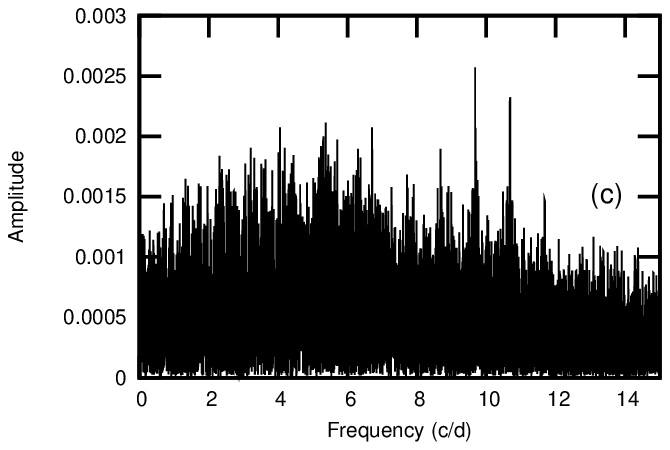}
\caption{Periodograms obtained by using all of the HUZ and CTE data. (a) The original periodogram with the observing window shown in the inset. (b) Prewhitened periodogram showing the peak at 2.93 c/d. (c)
Prewhitened periodogram showing the peak at 9.68 c/d. \label{periodograms}}
\end{figure}

\begin{figure}
\includegraphics[scale=1]{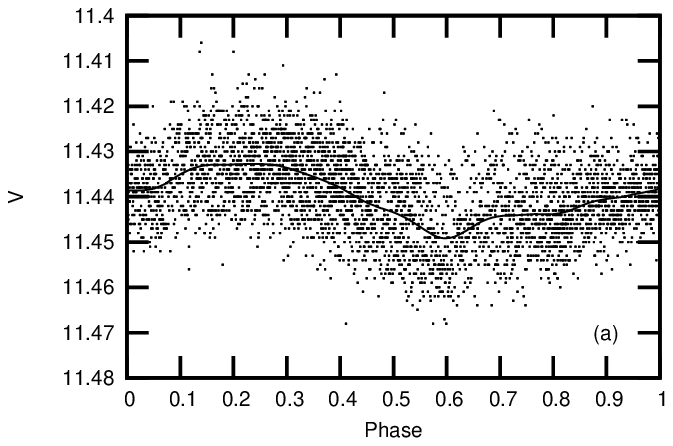}
\includegraphics[scale=1]{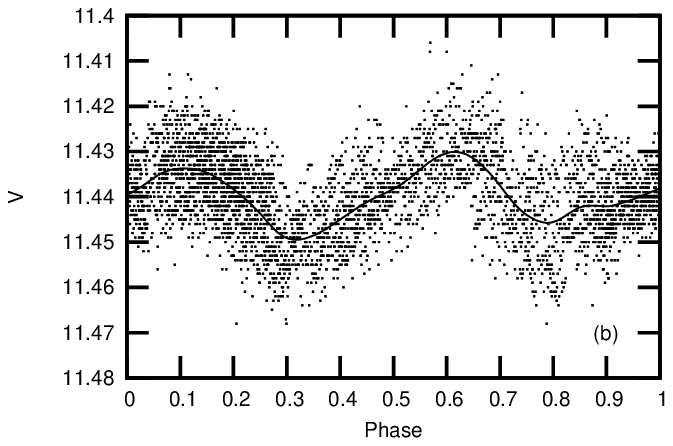}
\caption{Photometric $V$ band data from CTE folded two frequencies. (a) Folded at 2.924 c/d. (b) Folded at 1.462 c/d. The lower frequency
appears to be composed of two peaks at slightly different amplitudes. \label{fig1.46}}
\end{figure}

Selected periodograms associated with the entire dataset (all three observing seasons) are shown in Fig.~\ref{periodograms}. For the entire dataset, both the weighted and unweighted analysis found a low frequency component with the highest peak with a period of almost exactly 840 days, i.e. the time-span. This corresponds exactly to the 
expected effect of the single mean brightness offset in the first half of the 
central chunk of data (second season).  The 1.49 cycles/day (c/d) frequency is roughly half (not significantly different from half) that of the prominent period at 2.92 c/d and folded phase graphs at those two frequencies produces plots (Fig.~\ref{fig1.46}) that are very similar to directly observed all night time series (see Figs.~\ref{fig1a} and \ref{fig1b}). The 0.18 and 1.21 c/d periods appear to be spurious because of the distorted shape of the folded curves, 
sampling window effects, a non-periodic mean brightness shift, etc. The whole spectrum above $\sim$5 c/d can be very well approximated with a $1/f$ noise which may have an astrophysical origin. The 9.679 c/d peak is, however, very narrow, very strong, and much higher than the 
noise in the surrounding area of the spectrum. A typical method for assessing the signal to noise (S/N) ratio for a peak in the 
spectrum is to calculate the ratio between its amplitude and the mean 
amplitude of the surrounding peaks with a S/N $>$ 3.5 to 4 being a commonly adopted 
limit of significance (very roughly representing, via the central limit theorem, a probability of arising by chance of $<$5\%). The 9.679 c/d peak is significant by this definition having 
a S/N in the range 4.5 to 7.6, depending on width of the frequency bins. The frequencies of 2.92 c/d (and/or
1.49 c/d) and 9.68 c/d are therefore judged to be real from the analysis of the whole dataset. There was not much difference between the weighted and 
unweighted analysis, which shows that the overall photometric quality (from observers HUZ and CTE) is 
quite homogeneous and there is not much effect from outliers.

Fourier analysis of the middle season (160 days) showed a low frequency artefact (0.0044 c/d, a period of 222 days), a sampling 
artefact (0.996 c/d), a scatter of power between 2.7 and 2.99 c/d and again the 
9.678 c/d peak. As with using the whole dataset, phasing the data with the doubled 0.68 d 
period produces a double humped light curve with slightly different amplitudes for each hump.

Fourier analysis of the the last 30 days of data, which is the best subset, produced the results shown in the last three columns of Table \ref{pertab}.
The 2.9 and 9.7 c/d frequencies show up in every analysis. The phase diagram at 2.9 (or 1.5) c/d shows
a non-sinusoidal shape with 
sharp changes which is very suggestive of geometric variations.
The 1.52 c/d frequency may be a pure subharmonic within the uncertainties of the analysis and the 9.678 
c/d peak is very strong.

Rotational modulation can cause double humps if, for example, the modulation is caused by fixed star spots.
So, if rotation is the cause, the 
formal frequency associated with the 2.9 c/d peak from the Fourier spectrum needs to be halved to obtain the true frequency. Also the 2.9 c/d folded light curve seems to be a bit 
different in shape when the whole dataset or different subsets are used. If this is 
due to rotation then shifts are possible
if there is differential rotation on the B-star and/or spots change their 
latitudes. 
The 9.678 c/d frequency is very stable and very coherent (perhaps suggestive of some kind 
of geometric changes) and it is only 6 to 6.5 times longer than the X-ray pulse period of 25 minutes.

\begin{table*}
\caption{Frequencies found from the iterative Fourier analysis of the HUZ and CTE data, listed in order of significance. Frequencies that were found in the third season data subset (the best one) plus at least one of the other two analyses are highlighted in bold face.\label{pertab}}
\begin{center}
\begin{tabular}{ccccccccccc}
\hline
\multicolumn{3}{c}{Whole dataset} & $\mid$ & \multicolumn{3}{c}{Second season} & $\mid$ & \multicolumn{3}{c}{Last 30 days} \\
Frequency &  Amplitude   & S/N      & $\mid$ & Frequency &  Amplitude   & S/N  & $\mid$ & Frequency &  Amplitude   & S/N  \\ 
   (c/d)      &      (mmag)     &            & $\mid$ &   (c/d)       &      (mmag)     &        & $\mid$ &    (c/d)      &      (mmag)     &         \\
   \hline
0.00119        & 28.5       &  40          & & 0.9962         & 12.8        &  21           & & {\bf 2.9142} & {\bf 5.9} &  {\bf 11} \\
{\bf 2.9238}  & {\bf 6.5} &  {\bf 8.1} & & 1.2128         & 4.5          &  7.0          & & 0.1502         & 5.6         &  10 \\
0.1846          & 5.2         &  7.4         & & 2.7201         & 7.3          &  9.0          & & {\bf 1.5231} & {\bf 5.2} &  {\bf 9.2} \\
1.2124          & 4.3         &  5.8         & & 0.4222         & 7.9          &  13           & & {\bf 9.6786} & {\bf 2.6} &  {\bf 6.5} \\
{\bf 1.4983}  & {\bf 4.0} &  {\bf 5.0} & & {\bf 2.9894} & {\bf 5.8}  &  {\bf 7.2}  & &                     &               &        \\
{\bf 9.6791}  & {\bf 2.8} &  {\bf 7.6} & & 5.4630         & 3.5          &  4.0          & &                     &               &        \\
                     &               &                & & {\bf 9.6784} & {\bf 3.2}  &  {\bf 6.5}  & &                     &               &        \\
\hline 
\end{tabular}
\end{center}
\end{table*}

\begin{figure}
\includegraphics[scale=1,angle=-90]{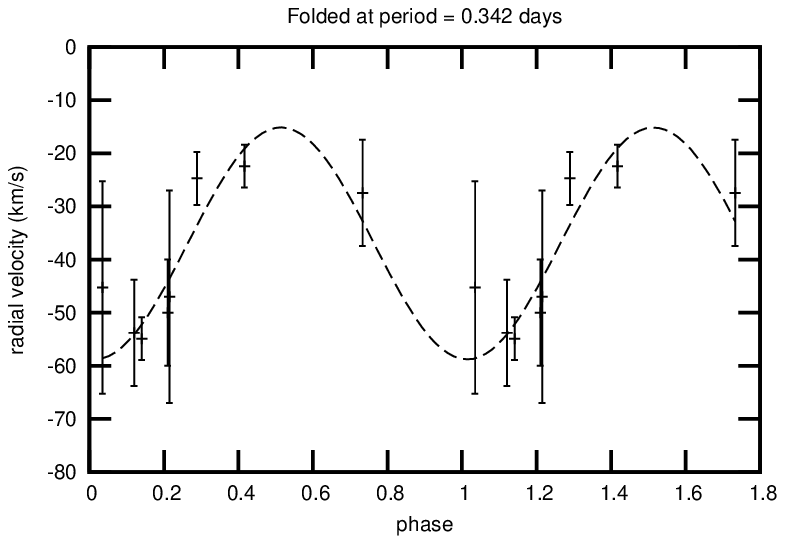}
\caption{Radial velocity data folded with a 0.342d period. 
The dashed line shows the best fitting sinusoid with a 0.342d period. 
\label{fig3}}
\end{figure}

\subsection{Radial Velocity Measurements}

Two approaches were taken to radial velocity determination: a cross correlation with a model spectrum as computed by \cite{munari2005} and direct line profile fitting for the cores of the four dominant absorption lines.

For the cross correlation approach, a high resolution (0.05 \AA/pix) template was used after being degraded to the resolution of the data. The template spectrum (shown at the bottom of Fig.~\ref{fig2}) was from the library of model spectra as computed
by \cite{munari2005}, with the following parameters: $T_{\mbox{\small eff}}=24000$K (see Table~\ref{apV831}), $\log g = 3.0$ (cgs units),
[M/H]=$-$0.5, $v_{\mbox{\small rot}}=0$ km s$^{-1}$, microturbulent velocity = 2 km s$^{-1}$, no [$\alpha$/Fe] enhancement,
old opacity distribution function (ODF) and no overshooting (see \citet{munari2005} for details).

The observed spectra were cross-correlated with
the template to compute relative radial velocities using the {\tt fxcor} task of IRAF. Barycentric corrections were applied using the IRAF task
{\tt rvcorrect}. For direct line profile fitting, a sum of Gaussian and Lorentzian functions was first tried but ultimately
a simple Gaussian fit of the line cores was used due to the limited S/N of the data. Of the four dominant lines in the spectra, the weakest line (Si II at $\lambda$4430) was dropped for all spectra and the remaining three lines were fitted. For the noisier spectra (due to clouds) the He~I $\lambda$4387 line could not be fitted, leaving only two fitted lines for those spectra. The resulting Doppler shifts were converted to heliocentric velocities using {\tt rvcorrect} and the average and standard deviation of the line velocities were computed for each spectrum to give an uncertainty measure.
The results (average line velocities) were comparable to the results obtained from cross-correlation. We considered the cross-correlation radial velocity values to be superior to the values obtained from individual line-fitting but the cross-correlation method gave no directly usable measures of error since the {\tt centre1d} peak-finding algorithm was used in IRAF. The {\tt centre1d} algorithm was determined to be better for finding the maximum of the cross-correlation function than function fitting methods because the cross-correlation function was frequently asymmetrical. So we used the standard deviation (SD) of the line-fits to quantify the error in the cross-correlation radial velocity values.
The results, excluding the three noisy spectra are plotted in Fig.~\ref{fig3}. The radial velocity values for the individual lines were averaged here but will be useful individually for a larger data set because each line potentially gives information about different layers in the Be star.

Large radial velocity changes were seen in the course of a single night so the data were folded
at the periods, and multiples of those periods, found in the period analysis of the photometry data. The radial velocity dataset is too small to permit any more extensive period searches.  After the radial velocities computed from the three obviously noisy spectra were discarded, the best results were found
at the prominent period of 0.342d. A sine curve was formally fit
to the data folded at that period using a nonlinear least-squares Marquardt-Levenberg algorithm. The
resulting fit, along with the measured radial velocity data are shown in Fig.~\ref{fig3}. For the sine curve fit, $\chi^{2}_{\mbox{\small red}}=1.07$ (5 degrees of freedom) where
$\chi^{2}_{\mbox{\small red}}$ is the reduced $\chi^{2}$ value obtained by weighting the terms in the residual sum of squares with the SD values from the line-fits. The systemic velocity from the fit was
$\gamma = -37.0 \pm 4.3$ km s$^{-1}$ and the radial velocity semi-amplitude was $K_{1}=21.8 \pm 3.6$ km s$^{-1}$. We should emphasise that this conclusion about the radial velocity period is uncertain and subsequent data may lead to support for a period other than the ones tried here.  

The value of $\gamma = -37.0 \pm 4.3$ km s$^{-1}$ closely matches the radial velocities for other stars in NGC 663 \citep{liu1989,liu1991}.  \citet{mermilliod2008} report a cluster mean radial velocity of $-33.09 \pm$0.34 km s$^{-1}$ with a dispersion of 1.74 km s$^{-1}$. Our measurements therefore provide further confirmation that the HMXB RX J0146.9+6121/LS I+61$^{\circ}$235 is a member of the open cluster NGC 663 
if we assume that the velocity kick from the supernova that formed the neutron star was non-existent or small. This is likely if RX J0146.9+6121/LS I+61$^{\circ}$235 belongs to the class of BeXs with long orbital periods \citep{pfahl2002} as we suspect for this system (see the Discussion). One possible mechanism for such a small or non-existent supernova velocity kick is an electron-capture collapse of a degenerate O-Ne-Mg core, of a lower mass progenitor, that can only happen in binary systems \citep{podsiadlowski2004}.

\section{Discussion}\label{sec4}

LS I+61$^{\circ}$235 (V831 Cas) was identified as the optical counterpart of RX J0146.9+6121
by \cite{motch1991}. They first identified the X-ray source from ROSAT data
and subsequently observed H$\alpha$ emission in the optical counterpart to confirm the BeX classification
of the source. 
RX J0146.9+6121 is a persistent low-luminosity BeX with low X-ray variability ($L_{\mbox{\tiny max}}/L_{\mbox{\tiny min}} \leq 10$) and it contains an X-ray pulsar with the longest pulse period known \citep{reig1999}. The pulsar period is
 variable around 1400s \citep{mereghetti2000} with 
variability
 presumably as the result of the increased mass transfer associated with occasional X-ray bursts. 

\cite{coe1993} placed the distance at 2.2 kpc
based on a reddening of $E(B-V)$ = 0.92 derived from {\em uvby} photometry (per the method of 
\cite{crawford1978}). This distance determination is consistent with LS I+61$^{\circ}$235 being a member of the open cluster NGC 663 \citep{pigulski2001} and was confirmed by \cite{reig1997} who further determined the astrophysical parameters listed in Table \ref{apV831}.
\cite{pigulski2001} give the age of NGC 663 to be 20--25 Myr. 
The age of NGC 663 has been difficult to determine because the cluster contains the highest known Be star abundance in the Galaxy, and Be stars tend to occupy anomalous positions in colour-magnitude diagrams \citep{fabregat2000}. \citet{tapia1991}, for example, give the lowest determination of 9 Myr. Using Str\"{o}mgren {\em uvby} colour indicies for non-Be stars in the cluster, \citet{fabregat2000} deduce an age of 23 Myr in agreement with the determination by \cite{pigulski2001}.
 \cite{coe1993} and \cite{reig1997,reig2000} obtained IR JHK photometry that showed long term variations of approximately 0.3 mag
for J (J$\sim$9.9), 0.3 mag for H (H$\sim$9.6) and up to 0.5 mag for K (K$\sim$9.4).

The Be decretion disk of LS I+61$^{\circ}$235 has been intensively studied through the investigation of Violet/Red (V/R) variations in the H$\alpha$ emission. The H$\alpha$, HeI $\lambda$6678 and the Paschen lines Pa11, $\lambda$8863, and Pa12, $\lambda$8750 all show variable emission \citep{reig2000}. Furthermore, the H$\alpha$ emission shows a quasi-cyclic variation of 1240 $\pm$ 30 days. The V/R variation of the H$\alpha$ line correlates with long-term changes in H$\alpha$ EW and IR (JHK) intensity. As shown by \cite{reig2000}, the V/R  and IR variations can be explained by the global one-armed oscillation model 
\citep{okazaki1991,okazaki1997,papaloizou1992} consisting of global $m=1$
oscillations of the cool Be disk in which an enhanced density perturbation develops on one side of
the disk and slowly precesses, in this case in the prograde direction, with one precession cycle being equal to the V/R cycle. The density
perturbation is confined to a few stellar radii in the disk and the precession period is expected
to be relatively insensitive to the disk size \citep{savonije1993}. Since
the variations are azimuthal, no variation in IR colour is expected and this is what has been observed.
Radial variation in the disk would cause IR colour variation. We will relate the motion of the one-armed density perturbation with the X-ray behaviour of RX J0146.9+6121 below.

\begin{table}
\caption{The astrophysical parameters of LS I+61$^{\circ}$235 \citep{reig1997}.\label{apV831}}
\begin{center}
\begin{tabular}{ll}
\hline
Spectral type & B1V \\
E($B-V$) & 0.93 $\pm$ 0.02 \\
$T_{\mbox{\tiny eff}}$ & 24000 $\pm$ 1500 K \\
Radius & 7 $\pm$ 1 R$_{\odot}$ \\
Mass & 11 $\pm$ 2 M$_{\odot}$ \\
$M_{V}$ & -3.1 $\pm$ 0.5 \\
Distance & 2.3 $\pm$ 0.5 kpc \\
BC & -2.4 $\pm$ 0.2 \\
$M_{\mbox{\tiny bol}}$ & -5.5 $\pm$ 0.5 ($1.2 \times 10^{4}$ L$_{\odot}$) \\
$\log g$ & 3.9 $\pm$ 0.2 $\log$(cm/s$^{2}$) \\
$v_{\mbox{\small rot}} \sin i$ & 200 $\pm$ 30 km s$^{-1}$ \\
\hline
\end{tabular}
\end{center}
\end{table}

The strongest results arising from our data for LS I+61$^{\circ}$235/RX J0146.9+6121 are three frequencies in the
$V$ band light curve: 2.9 c/d, 1.5 c/d and 9.7 c/d. That is, the periods 0.34d, 0.67d and 0.10d are present. Such short periods have been argued for other Be stars to be due to rotation or non-radial pulsation
\citep{balona1990}.  The 0.67d period is almost surely a subharmonic of the 0.34d period and, if the light
variation is caused by a geometrical effect such as ellipsoidal variation (the Z-shape of the folded light curve implies a more complex geometry than simple ellipsoidal variation), then the spin period of the Be star is
0.68d. This conclusion can be checked using the astrophysical parameters for LS I+61$^{\circ}$235 previously published
by \cite{reig1997} (see Table \ref{apV831}). Using the value of 2.9238 c/d (0.34d period) as determined from the analysis of all three observing seasons and a Be star radius of 7$\pm$1 R$_{\odot}$ we find a surface velocity
of 1036
$\pm$148
km s$^{-1}$. With a rotation frequency of half that, a surface velocity of 518
$\pm$74
 km s$^{-1}$ results. The break-up velocity is
\begin{equation}\label{vcrit}
v_{\mbox{\small crit}} = \sqrt{\frac{G M_{*}}{R_{*}}}.
\end{equation}
Using $M_{*} = 11$ M$_{\odot}$ and $R_{*} = 7$ R$_{\odot}$, the critical velocity is $v_{\mbox{\small crit}} =
547$ km s$^{-1}$. So the rotation period cannot be 0.34d or the star would break-up. A rotation period of
0.68d is near-critical, a condition that is believed to hold for Be stars 
(e.g. from statistical studies like that of \cite{chauville2001}, from the direct interferometric measurement of the oblateness of the Be star Achernar by \cite{domiciano2003} and,
from theoretical considerations on how the decretion disk is formed like that of \cite{lee1991}).
So it is possible that the rotation period may be
0.68d with the 0.34d period representing when opposite sides of the star are presented to us. 

With a rotation period of 0.67d and $v_{\mbox{\small rot}} = 518$ km s$^{-1}$ we can use the previously determined value of $v_{\mbox{\small rot}} \sin i = 200$ km s$^{-1}$ to deduce that the inclination is
$i = 23^{\circ}$. Taking into account the error reported for $R_{*}$ and $v_{\mbox{\small rot}} \sin i$ we
have $i = 23\ ^{+10}_{ -6}$ degrees.

Next we investigate the possibility that the 0.34d period is the binary orbital period since we have some evidence that the radial velocity period is 0.34d.
The mass function, $f(M)$, with masses $M_{1}$, the mass of the Be star and $M_{2}$, the mass of the compact companion, given in solar masses, the orbital period $P_{\mbox{\small orb}}$ in days and the semi-amplitude
of the radial velocity variation, $K_{1}$, given in km s$^{-1}$, is
\begin{equation}
f(M) = \frac{M_{2}^{3} \sin i}{(M_{1} + M_{2})^{2}} = (1.0361\! \times\! 10^{-7})(1 - e^{2})^{\frac{3}{2}} K_{1}
P_{\mbox{\small orb}}\label{massfn}
\end{equation}
where $e$ is the eccentricity of the orbit \citep{hilditch2001}. Taking $e=0$, $M_{1}=11$ M$_{\odot}$, $K_{1} = 22$ km s$^{-1}$ and 
$i=23^{\circ}$ (assuming the Be spin axis and the orbital axis to be aligned) in Eq.~\ref{massfn} gives the mass of the compact companion as $M_{2}=0.45$ M$_{\odot}$ for $P_{\mbox{\small orb}}
= 0.34$d. The semi-major axis of the orbit is given by Kepler's law as
\begin{equation}\label{asep}
a = \left( \frac{P_{\mbox{\small orb}}^{2}}{4 \pi^{2}}G(M_{1}+M_{2}) \right)^{\frac{1}{3}}.
\end{equation}
Using $M_{2}=0.45$ M$_{\odot}$ and $P_{\mbox{\small orb}}= 0.34$d gives $a = 4.6$ R$_{\odot}$. Since the radius of the Be star is 7$\pm$1 M$_{\odot}$, the proposed
orbit is clearly too small, so 0.34d cannot be the orbital period. 

The mass and luminosity of LS I+61$^{\circ}$235 as given in Table~\ref{apV831} put it in the middle of the $\beta$ Cep instability strip on the Hertzsprung-Russell diagram \citep{pamyatnykh1999,stankov2005}. In a recent catalogue \citep{stankov2005}, $\beta$ Cep stars are defined to be massive nonsupergiant variable stars with spectral type O or B whose light, radial velocity and/or line profile variations are caused by low-order pressure and gravity mode pulsations. For low order radial modes the light curve and radial velocity will have the same period. This is what appears here if we take the radial velocity period to be 0.34d. In that case the pulsation constant, $Q$, given by
\begin{equation}
Q = P \sqrt{\frac{3 M}{4 \pi R^{3}}}
\end{equation}
where $P$ is the period and $M$ and $R$ are the mass and radius of the star in solar units, is 0.03$\pm$0.01d, if the astrophysical parameters from Table~\ref{apV831} are used. The modal $Q$ of known $\beta$ Cep variables is 0.033d \citep{stankov2005} which corresponds to the radial fundamental mode. 

If, between the two periods 0.34d and 0.67d, we take 0.34d as the pulsation period, we can not interpret the
0.67d period as a harmonic due to a non-sinusoidal pulsation in time. Thus the interpretation of the 0.67d period
as the Be star spin period with a spin axis inclination of 23$^\circ$ may still hold. The combination of 0.34d being a harmonic (or close to a harmonic) of the 0.67d period plus a pulsation period of 0.34d would make the 0.34d period the prominent one, as observed, since two processes would contributing to the light variation at that period. A simpler interpretation is that the 0.34 radial velocity period is spurious and the 0.34d photometric period is simply a harmonic of the 0.67d period due to the non-sinusoidal shape of the 0.67d period. Further radial velocity data can resolve this ambiguity. If the 0.10d period is due to pulsation and 0.34d is the fundamental mode, then it must be a higher order p mode.

Finally, we speculate on the origin of the ``super'' X-ray outbursts spaced at approximately 450d. The prograde period of the one-armed density enhancement in the Be disk has been measured to be approximately 1240d with some variability \citep{reig2000}. If we assume that the compact star also orbits in a prograde direction (relative to the Be star spin) then we might expect a super outburst to occur when the arm and the compact star line up. Given the arm period of 1240d and an outburst (alignment) period of 450d, this means that the compact star's orbital period is 330d in agreement with an expectation of about 300d from the Corbet spin-orbit relation \citep{corbet1986,reig2007}. This conclusion is roughly consistent with the 184d period found in the {\it RXTE} ASM data which would be explained by the occurrence of X-ray bursts when the compact star passed through the plane of the decretion disk. The cause of the apparent 90d intervals between the minor outbursts, if these intervals are significant, is, however, not clear in this interpretation since the minor outbursts would have to occur four times per orbit. Perhaps, if the orbit is sufficiently inclined to the Be disk, the mass flow onto the neutron star will switch poles 90$^{\circ}$ in the orbit from the disk crossing nodes and this switch, through some as yet undetermined mechanism, could lead to a minor X-ray outburst.

An orbital period of 330d matches expectations from the Corbet spin-orbit relation but is it consistent with the X-ray luminosity of RX J0146.9+6121 and typical mass loss rates for Be stars? Based on a distance of 2.5 kpc, \cite{haberl1998} determined that the X-ray luminosity, $L_{x}$ varies between $2 \times 10^{34}$ and $2 \times 10^{35}$ erg s$^{-1}$. The X-ray lumiosity
is related to the mass accretion rate, $\dot{M}_{2}$ onto the neutron star through the conversion of gravitational potential energy to luminosity by
\begin{equation}
\dot{M}_{2} = \frac{L_{x}R_{2}}{G M_{2}}
\end{equation}
where $G$ is the gravitational constant and where we may take $R_{2} = 10^{6}$ cm and
$M_{2} = 1.4$ M$_{\odot}$ to be the radius and mass of the neutron star. So $\dot{M}_{2}$ is between $2 \times 10^{-12}$ and
$2 \times 10^{-11}$ M$_{\odot}$ yr$^{-1}$. Following \cite{frank2002} we assume that mass leaves the Be star at escape velocity and will be captured within a cylindrical region that has an axis through the neutron star in the direction of the relative wind velocity. Defining $v_{w}$ to be the wind velocity at a large distance from the Be star and $v_{\rm rel}$ as the wind velocity relative to the neutron star, we have $v_{w} =\sqrt{2} v_{\rm crit}$ (the escape velocity, see Eq.~\ref{vcrit} for the definition of $v_{\rm crit}$) and $v_{\rm rel} = v_{w}$ since the Keplerian velocity about the Be star at the neutron star distance, $a$, is small relative to $v_{w}$. So the cylindrical capture axis direction would be essentially towards the Be star. The radius of the cylindrical capture region, $r_{\rm acc}$, is where the kinetic energy of the wind equals the gravitational potential of the neutron star \citep{bondi1952},
\begin{equation}
r_{acc} = \frac{2 G M_{2}}{v_{\rm rel}^{2}}.
\end{equation}
The ratio of the emitted wind, $-\dot{M}_{1}$, to the captured wind, $\dot{M}_{2}$, mass flow rates is therefore
\begin{equation}
\frac{\dot{M}_{2}}{-\dot{M}_{1}} = \frac{\pi r_{\rm acc}^{2}}{4 \pi a^{2}}
\end{equation}
where the binary separation, $a$, is given by Eq.~\ref{asep}. If $v_{\rm rel} = \sqrt{2} v_{\rm crit}$ then
\begin{equation}\label{redfrac}
\frac{\dot{M}_{2}}{-\dot{M}_{1}} = \frac{1}{4} \left( \frac{M_{2}}{M_{1}} \right)^{2}
\left( \frac{R_{1}}{a} \right)^{2}.
\end{equation}
For $M_{1} = 11$ M$_{\odot}$, $M_{2} = 1.4$ M$_{\odot}$ and an
orbital period of 330d, $a = 480$ R$_{\odot}$, and using $R_{1} = 7$ R$_{\odot}$ in Eq.~\ref{redfrac} gives a ratio of emitted to captured wind of $10^{-6}$. So the inferred Be wind rate, $-\dot{M}_{1}$, would be between $2 \times 10^{-6}$ and
$2 \times 10^{-5}$ M$_{\odot}$ yr$^{-1}$. This is consistent with Be mass outflow rates of $\sim$10$^{-6}$ M$_{\odot}$ yr$^{-1}$ seen observationally and expected on
theoretical grounds \citep{lamers2000,vink2000}. However, we should also note that some observational and theoretical studies have come to the conclusion that the Be mass loss rate is much lower at $10^{-9}$ to $10^{-8}$ $M_{\odot}$ yr$^{-1}$ 
\citep{waters1988,haberl1998}.

\section{Conclusion}\label{sec5}

The $V$ band light curve of LS I+61$^{\circ}$235 shows three strong periodicities at 0.34d, 0.67d and 0.10d. In our interpretation, the 0.34d period may be from pulsation in the radial fundamental mode, the 0.68d period is the Be star spin period and the 0.10d period is a higher order p mode pulsation.  The
spin period, when combined with a previous determination of the astrophysical parameters of the Be star,
gives an inclination for the Be spin axis of $i = 23\ ^{+10}_{ -6}$ degrees. Comparing the latest catalogue of $\beta$ Cep variables \citep{stankov2005} to the latest catalogue of Galactic HMXBs \citep{liu2006} reveals that LS I+61$^{\circ}$235 would be the first confirmed $\beta$ Cep member of a HMXB if subsequent observations uphold these conclusions. The systemic radial velocity confirms that LS I+61$^{\circ}$235 is a member of the cluster NGC 663.
If we interpret that the X-ray ``super'' outbursts happen when the compact star lines up with the one-armed Be decretion disk enhancement, then the orbital period is approximately 330d.  This orbital period agrees with the value expected from the Corbet spin-orbit relation \citep{corbet1986,reig2007}.

The variability of the $I_{C}$ light curve looks to be similar to the variability of the $V$ band light
curve and there may be some variability in $B-V$ colour. More observations are needed to investigate these
possible colour variations which should exist if pulsation is present. 

\section*{Acknowledgements}

GES is supported by a discovery grant 
from the Natural Sciences and Engineering Research Council of Canada (NSERC).
Thanks to R. Sordo for providing the template spectra. Thanks to S.C.~Orlando for his two photometric data points and a discussion of comparison star variability. Many thanks to Arne Henden for the reduced SRO data
used to define the comparison stars. The assistance of AAVSO staff members Aaron Price, Matt Templeton and Elizabeth Wagner in coordinating alert notices and accommodating the HMXB observing project are greatly appreciated. Thanks to Dmitry Monin for observing support with the DAO telescope. The Digitized Sky Surveys used to produce Fig.~\ref{figcomps} were produced at the Space Telescope Science Institute under U.S. Government grant NAG W-2166. The images of these surveys are based on photographic data obtained using the Oschin Schmidt Telescope on Palomar Mountain and the UK Schmidt Telescope. The plates were processed into the present compressed digital form with the permission of these institutions. The 30-cm telescopes used by observer HUZ were provided by the Department of Physics and Engineering Physics at the University of Saskatchewan and are primarily maintained by Stan Shadick.


\label{lastpage}

\end{document}